\documentclass[twocolumn,aps,showpacs]{revtex4}
\usepackage{amssymb}
\usepackage{epsf}
\usepackage{amsmath}
\usepackage{graphicx}

\def\lsim{\lower -0.3ex \hbox{$<$} \kern -0.75em \lower 0.7ex \hbox{$\sim$}}
\def\gsim{\lower -0.3ex \hbox{$>$} \kern -0.75em \lower 0.7ex \hbox{$\sim$}}

\begin{document}

\title{Spin-orbit coupling and broken spin degeneracy in multilayer graphene}
\author{Edward McCann$^{1}$ and Mikito Koshino$^{2}$}
\affiliation{
$^{1}$Department of Physics, Lancaster University, Lancaster, LA1
4YB, UK\\
$^{2}$Department of Physics, Tokyo Institute of
Technology, 2-12-1 Ookayama, Meguro-ku, Tokyo 152-8551, Japan}

\begin{abstract}
Since the lattices of $ABA$-stacked graphene multilayers with an even number of layers,
as well as that of monolayer graphene, satisfy
spatial-inversion symmetry, their electronic bands must be spin degenerate
in the presence of time-inversion symmetry.
In intrinsic monolayer and bilayer graphene,
 when symmetry is not broken by external fields,
 the only spin-orbit coupling present at low energy near the corner of
the Brillouin zone is the Kane-Mele term, that opens a bulk energy gap
but does not break the spin degeneracy of the energy bands
[C. L. Kane and E. J. Mele, Phys. Rev. Lett. \textbf{95}, 226801 (2005)].
However, spin splitting is allowed in
multilayers with an odd number of layers ($\geq 3$) because their
lattices do not satisfy spatial inversion symmetry.
We show that, in trilayer graphene, in addition to the Kane-Mele term,
there is a second type of intrinsic spin-orbit coupling present at
low energy near the corner of the Brillouin zone.
It introduces a Zeeman-like spin splitting of the energy bands
at each valley, with an opposite sign of the effective magnetic field in the
two valleys. We estimate the magnitude of the effective field to be
$\sim 2$T.
\end{abstract}


\maketitle

Interest in graphene \cite{novo04,castroneto} stems partly from
the presence of two Dirac cones in the low-energy electronic band
structure, each of them supporting chiral quasiparticles.
The origin of the Dirac-like spectrum lies in the fact
that the honeycomb lattice of graphene has two inequivalent atomic sites,
called $A$ and $B$, and the amplitude of the electronic
wave function on them acts as an additional degree of freedom
known as pseudospin. An `up' component of pseudospin pointing
perpendicular to the graphene plane would correspond to electronic
density solely on the $A$ sublattice sites, whereas `down' pseudospin
corresponds to density on the $B$ sublattice. In practice, electronic
density is usually shared equally between the sublattices so that
the pseudospin is a linear combination of `up' and `down', and it lies
in the plane of the graphene sheet. There are two Dirac cones,
centered at inequivalent corners of the Brillouin zone which are
denoted $K_{+}$ and $K_{-}$ and are also referred to as valleys.
The valleys introduce another spin-like degree of freedom into the electronic Hamiltonian.

The band structure described above is modified
when the spin of the electron is taken into account.
Kane and Mele \cite{kanemele05} introduced a spin-orbit
coupling term that exists at the center of each valley,
\begin{eqnarray}
\mathcal{H}_{KM} &=& \alpha \,\! \Pi_z \sigma_z S_z \, , \label{hkm}
\end{eqnarray}
where $\alpha$ is a parameter and Pauli matrices $\Pi_z$, $\sigma_z$, $S_z$
act in $K_{+}$/$K_{-}$ valley, $A$/$B$ sublattice, and $\uparrow$/$\downarrow$ spin space, respectively.
It satisfies time and spatial inversion symmetry requirements by coupling
the out-of-plane component of electronic spin with the
out-of-plane component of pseudospin.
This does not break the spin and valley degeneracy of the energy bands,
but it changes the balance between the $A$ and $B$
sublattices, opening a band gap and
realizing a new, topological state of matter, a quantum
spin Hall insulator \cite{kanemele05}.

In this paper, we explain how time and spatial-inversion symmetry
influence spin and valley degeneracy in $ABA$-stacked (Bernal)
multilayer graphene composed of $N$ layers. In the presence of time-reversal symmetry, when
the atomic lattice satisfies spatial inversion symmetry, and such symmetry
is not broken by external fields,
the electronic bands must be spin degenerate. This holds
for monolayer and bilayer graphene where the only intrinsic spin-orbit coupling
term \cite{kanemele05,min06,dhh06,vgeld09,guinea10,liu10} present at low-energy near
the center of the valley is of the Kane-Mele type, Eq.~(\ref{hkm}).
However, the lattices of multilayers with odd $N$ ($N \geq 3$)
do not satisfy spatial inversion symmetry so that
spin splitting of the energy bands is allowed.
We show that, in trilayer graphene, in addition to the Kane-Mele term, Eq.~(\ref{hkm}),
there is a second type of intrinsic spin-orbit coupling present at
low-energy near the center of the valley,
\begin{eqnarray}
\mathcal{H}_{ABA} &=& \beta \,\! \Pi_z S_z \, . \label{haba}
\end{eqnarray}
It couples the out-of-plane component of electronic spin with the
out-of-plane component of the valley `spin' degree of freedom.
This introduces a Zeeman-like spin splitting of the energy bands
at each valley, of magnitude proportional to the parameter $\beta$,
with an opposite sign of the effective Zeeman field in the
two valleys.

\begin{figure}[t]
\centerline{\epsfxsize=0.9\hsize \epsffile{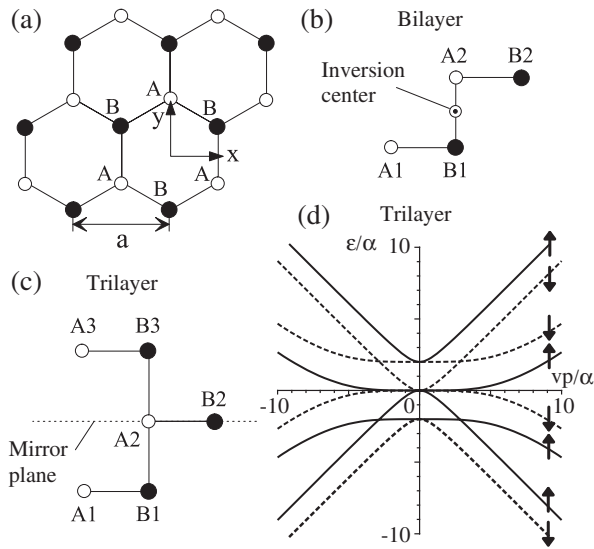}}
\caption{(a) Plan view of the lattice of a single layer of graphene showing inequivalent
$A$ and $B$ sites.
(b) Schematic side view of the unit cell of bilayer graphene and (c) of Bernal-stacked
trilayer graphene, showing the inversion center for bilayer
and the mirror plane for trilayer. (d) Sketch of the trilayer energy bands in the vicinity of
the valley center, using Eqs.~(\ref{em}) and (\ref{eb}), where
we use $a_m = a_b = b_m = b_b = 0$,
$\alpha_m = \alpha_b = \beta_m = - \beta_b \equiv \alpha$
with the value $\alpha = \gamma_1 / 20$ chosen for illustrative purposes.
}
\label{fig:1}
\end{figure}

We begin by briefly describing how time and spatial symmetries
impose constraints on the spectra of graphene multilayers,
before discussing the particular examples of bilayer
and trilayer graphene. We consider intrinsic spin-orbit terms
that exist in samples in the absence of external electric or magnetic fields.
For example, we neglect the influence of a transverse electric field that
breaks reflection symmetry and produces an additional,
Rashba-like term \cite{kanemele05,rashba,min06,dhh06}.
Time reversal symmetry relates energy bands with opposite spin components $\{ \uparrow , \downarrow \}$
and momenta $\mathbf{k}$ and $-\mathbf{k}$:
$\epsilon_{\uparrow} ( \mathbf{k} ) = \epsilon_{\downarrow} ( -\mathbf{k} )$.
In monolayer graphene \cite{kanemele05}, and in multilayers with even $N$,
including bilayers \cite{novo06,mcc06a,ohta06,guinea06},
the lattice, Fig.~\ref{fig:1}(a) and (b), obeys spatial inversion symmetry
$(x,y,z) \rightarrow (- x, -y , -z)$,
which relates energy bands with the same spin component
(as spin is an axial vector) and opposite momenta:
$\epsilon_{\uparrow} ( \mathbf{k} ) = \epsilon_{\uparrow} ( -\mathbf{k} )$.
Combining time and spatial inversion symmetry guarantees spin
degeneracy of the energy bands,
$\epsilon_{\uparrow} ( \mathbf{k} ) = \epsilon_{\downarrow} ( \mathbf{k} )$,
even in the presence of finite spin-orbit coupling.
The situation is different in multilayers with odd $N$, starting
from trilayer graphene ($N=3$), Fig.~\ref{fig:1}(c), because their lattices do not
satisfy spatial inversion symmetry \cite{latil06,manes07,kosh09a}.
When time reversal symmetry holds, the energy of bands with
opposite spin and momenta must still be equal,
$\epsilon_{\uparrow} ( \mathbf{k} ) = \epsilon_{\downarrow} ( -\mathbf{k} )$,
but spin splitting of the energy bands at the same momentum,
$\epsilon_{\uparrow} ( \mathbf{k} ) \neq \epsilon_{\downarrow} ( \mathbf{k} )$,
is allowed, Fig.~\ref{fig:1}(d).

{\em Bilayer graphene:}
In multilayers with an even number of layers,
the lattice is symmetric with respect to spatial inversion
symmetry $P$ [$(x , y , z) \rightarrow (-x , -y , -z)$],
because the point group of the lattice $D_{3d}$
\cite{latil06,manes07}
($\{ E, 2C_3, 3C_2^{\prime}, i, 2S_6, 3\sigma_d\}$)
can be regarded as a direct product of group
$D_3$ ($\{ E, 2C_3, 3C_2^{\prime}\}$)
with the inversion group $C_i$ ($\{ E, i \}$).
Here, we focus on the particular case of bilayer
graphene \cite{novo06,mcc06a,ohta06,guinea06}, which consists of two
coupled layers of carbon atoms arranged
on a honeycomb lattice with inequivalent sites $\{ A1 , B1 \}$
and $\{ A2 , B2 \}$ on the bottom and top layers,
respectively, Fig.~\ref{fig:1}(b). The low-energy tight-binding model takes into
account a single $p_z$ orbital per site, with parameter $\gamma_0$
taking into account nearest-neighbor hopping, $A1$-$B1$
and $A2$-$B2$, within each layer.
The layers are arranged so that sites
$B1$ and $A2$ are directly below and above each other,
connected by interlayer coupling $\gamma_1 \sim 0.4$eV.
Then, the band structure consists of four bands. Two of them,
corresponding to the coupled orbitals on the
$B1$ and $A2$ sites, are split away from zero energy by
$\pm \gamma_1$. The other two bands,
formed by effective hopping between $A1$ and $B2$ sites, are parabolic and
touching near zero energy.

The parabolic low-energy bands may be described
using a basis of electronic wave functions
with components $\psi_{i,\xi,s}$
where $i = \{ A1 , B2 \}$ labels the
relevant lattice sites,
$s = \{ \uparrow , \downarrow \}$ spin components,
and $\xi = \pm 1$ distinguishes
between the valleys located at
$\mathbf{K}_{\xi }=\xi ({\textstyle\frac{4}{3}}\pi a^{-1},0)$
with lattice constant $a$.
Neglecting spin, the low-energy bands are approximately described
by the effective Hamiltonian \cite{mcc06a}
$\mathcal{H}_{2}^{(0)} = - (v^2/\gamma_1) S_0 \left(
\Pi_0 \sigma_x \left[ p_x^2 - p_y^2 \right] +
2 \Pi_z \sigma_y p_x p_y \right)$,
where we use direct products of Pauli matrices
$\sigma_{x,y,z},\sigma_{0}\equiv \hat{1}$ acting in the $A1$/$B2$ sublattice space,
$\Pi_{x,y,z},\Pi_{0}\equiv \hat{1}$ acting in the valley space,
and $S_{x,y,z},S_{0}\equiv \hat{1}$ acting in the spin space.
Here, $v=\left( \sqrt{3}/2\right) a\gamma_{0}/\hbar$
is the monolayer Fermi velocity
and $\mathbf{p} = (p_x , p_y)$ is the momentum
measured with respect to the center of the
valley, $\mathbf{k} = \hbar \mathbf{K}_{\xi } + \mathbf{p}$.

We begin by exploring which spin-orbit terms, if any,
satisfy time and space symmetries of the lattice.
Apart from symmetry, we assume that low-energy states
are composed from orbitals on the $A1$, $B2$ sites,
so that the following argument does not rely on details
of a particular model.
In the above basis, the time reversal operator is
$T = \Pi_x \sigma_0 S_y {\cal K}$ where ${\cal K}$
represents complex conjugation,
and translation by a distance equal to the lattice constant $a$ along the
$x$-axis is given by $e^{i4\pi \Pi_z/3}$.
Generators of the point group
of the lattice $D_{3d}$ are an active rotation anti-clockwise
by angle $2\pi /3$, $C_3 = - e^{i \pi \Pi_z \sigma_z/3}e^{-i\pi S_z/3}$,
rotation by angle $\pi$ about the $x$ axis,
$C_2^{\prime} = -i \Pi_0 \sigma_x S_x$,
and spatial inversion, $i = \Pi_x \sigma_x S_0$.
We focus on the center of the valley, $p = |\mathbf{p}| = 0$,
and consider all possible terms describing coupling between valley,
sublattice and spin of the form
$\Pi_i \sigma_j S_k$ where $i = \{0,x,y,z \}$, $j = \{0,x,y,z \}$,
$k = \{0,x,y,z \}$. Then, we determine which of them
is invariant with respect to the time and space symmetries
of the bilayer lattice.
Except for a constant term $\Pi_0\sigma_0S_0$,
there is only one,
which is a term of the Kane-Mele type, Eq.~(\ref{hkm}), \cite{kanemele05,min06,dhh06,vgeld09,guinea10,liu10}.
As in a monolayer \cite{kanemele05}, it leads to
the opening of a gap at the center of the valley,
but doesn't break spin and valley degeneracy.

{\em Trilayer graphene: }
In odd-$N$ multilayers with $N\geq 3$, the point group
$D_{3h}$ ($\{ E, 2C_3, 3C_2^{\prime}, \sigma_h, 2S_3, 3\sigma_v\}$)
\cite{latil06,manes07} can be regarded as a direct product of the group
$D_3$ ($\{ E, 2C_3, 3C_2^{\prime}\}$) with the reflection group $C_s$ ($\{ E, \sigma_h \}$).
With respect to the even-$N$ multilayers, spatial inversion
is replaced by mirror reflection $\sigma_h$
[$(x , y , z) \rightarrow (x , y , -z)$].
Trilayer graphene
\cite{lu06,guinea06,latil06,part06,kosh_mlg,aoki07,kosh09a,crac08,avet}
consists of three
coupled layers of carbon atoms arranged
on a honeycomb lattice with inequivalent sites $\{ A1 , B1 \}$,
$\{ A2 , B2 \}$ and $\{ A3 , B3 \}$ on the bottom, middle and top layers,
respectively, Fig.~\ref{fig:1}(c). The simple tight-binding model using a single
$p_z$ orbital per atom includes coupling $\gamma_0$
between nearest-neighbor sites within each layer, $A1$-$B1$, $A2$-$B2$
and $A3$-$B3$, as well as interlayer coupling $\gamma_1$ between sites
$B1$ and $A2$, and $A2$ and $B3$, that are directly below and above each other.

Linear combinations of atomic orbitals may be separated \cite{kosh_mlg,kosh09a} into a pair
that are odd with respect to mirror reflection symmetry $\sigma_h$,
$\phi_{1,\xi,s}^{(m)} = [\psi_{A1,\xi,s} - \psi_{A3,\xi,s}]/\sqrt{2}$,
$\phi_{2,\xi,s}^{(m)} = [\psi_{B1,\xi,s} - \psi_{B3,\xi,s}]/\sqrt{2}$,
and four that are even with respect to $\sigma_h$,
$\phi_{1,\xi,s}^{(b)} = [\psi_{A1,\xi,s} + \psi_{A3,\xi,s}]/\sqrt{2}$,
$\phi_{2,\xi,s}^{(b)} = \psi_{B2,\xi,s}$,
$\phi_{3,\xi,s}^{(b)} = \psi_{A2,\xi,s}$,
$\phi_{4,\xi,s}^{(b)} = [\psi_{B1,\xi,s} + \psi_{B3,\xi,s}]/\sqrt{2}$.
Written in a basis of states $\phi_{i,\xi,s}^{(m/b)}$, the Hamiltonian of trilayer
graphene separates into two parts \cite{kosh_mlg,kosh09a}.
The first has a basis of odd orbitals $\phi_{1,\xi,s}^{(m)}$, $\phi_{2,\xi,s}^{(m)}$,
and it has a form analogous to that of monolayer graphene,
\begin{eqnarray}
\mathcal{H}_{m}^{(0)} &=& v S_0 (
\Pi_z \sigma_x^{(m)} p_x +
\Pi_0 \sigma_y^{(m)} p_y ) \, ,
\label{hm0}
\end{eqnarray}
where matrices
$\sigma_{0,x,y,z}^{(m)}$ act in
the $\phi_1^{(m)},\phi_2^{(m)}$ sublattice space.
The second part of the Hamiltonian has a basis of four even orbitals
$\phi_{i,\xi,s}^{(b)}$, resulting in four parabolic bilayerlike bands.
As in a bilayer, two bands touch near zero energy. They are related to
orbitals $\phi_{1,\xi,s}^{(b)}$, $\phi_{2,\xi,s}^{(b)}$
and may be described by an effective quadratic Hamiltonian
\begin{eqnarray}
\!\!\!\!\!\!\mathcal{H}_{b}^{(0)} &=& - \frac{v^2 S_0}{\sqrt{2}\gamma_1} [
\Pi_0 \sigma_x^{(b)} ( p_x^2 - p_y^2 ) +
2 \Pi_z \sigma_y^{(b)} p_x p_y ]  ,
\label{hb0}
\end{eqnarray}
where matrices $\sigma_{0,x,y,z}^{(b)}$ act in
the $\phi_1^{(b)},\phi_2^{(b)}$ sublattice space.

An analysis of the transformation properties of the trilayer
Hamiltonian under symmetries of the lattice is relatively simple in
the basis of orbitals $\phi_{i,\xi,s}^{(m/b)}$ because
time and spatial symmetry operators do not mix the monolayerlike
and bilayerlike parts of the Hamiltonian. We therefore consider them separately.
In the monolayerlike part, the time reversal operator is
$T = \Pi_x \sigma_0^{(m)} S_y {\cal K}$,
and translation by a distance equal to the lattice constant $a$ along the
$x$-axis is given by $e^{i4\pi \Pi_z/3}$.
Generators of the group
$D_{3h}$ are an active rotation anti-clockwise
by angle $2\pi /3$,
$C_3 = e^{-i\pi S_z/3}
[ ( 1 - \sigma_z^{(m)} )/2 + e^{-i 2\pi\Pi_z/3} ( 1 + \sigma_z^{(m)} )/2]$,
rotation by angle $\pi$ about the $y$ axis,
$C_2^{\prime} = i \Pi_x \sigma_0^{(m)} S_y$,
and mirror reflection symmetry, $\sigma_h = i \Pi_0 \sigma_0^{(m)} S_z$.

We focus on the center of the valley, $p = 0$,
and consider all possible terms describing coupling between valley,
sublattice and spin of the form
$\Pi_i \sigma_j S_k$ where $i = \{0,x,y,z \}$, $j = \{0,x,y,z \}$,
$k = \{0,x,y,z \}$. Then, we determine which of them
is invariant with respect to the time and space symmetries of the lattice:
\begin{eqnarray}
\mathcal{H}_{m}^{(1)} &=& a_m \Pi_0 \sigma_0^{(m)} S_0 + b_m \Pi_0 \sigma_z^{(m)} S_0 \nonumber\\
&& \, + \, \alpha_{m} \Pi_z \sigma_z^{(m)} S_z
+ \beta_{m} \Pi_z \sigma_0^{(m)} S_z \, , \label{hmso}
\end{eqnarray}
The first line of Eq.~(\ref{hmso}) contains spin-independent terms,
the second line contains spin-orbit terms.
We repeat the symmetry analysis for the bilayerlike
part of the Hamiltonian.
We consider the
two low-energy bands that touch near zero energy, related to
orbitals $\phi_{1,\xi,s}^{(b)}$, $\phi_{2,\xi,s}^{(b)}$.
For that basis, $T = \Pi_x \sigma_0^{(b)} S_y {\cal K}$,
translation is $e^{i4\pi \Pi_z/3}$,
$C_3 = e^{-i\pi S_z/3} e^{-i 2\pi\Pi_z\sigma_z^{(b)}/3}$,
$C_2^{\prime} = i \Pi_x \sigma_z^{(b)} S_y$,
and $\sigma_h = - i \Pi_0 \sigma_0^{(b)} S_z$.
The invariant terms at the center of the valley are
\begin{eqnarray}
\mathcal{H}_{b}^{(1)} &=& a_b \Pi_0 \sigma_0^{(b)} S_0 + b_b \Pi_0 \sigma_z^{(b)} S_0
\nonumber \\
&&  \, + \, \alpha_{b} \Pi_z \sigma_z^{(b)} S_z
+ \beta_{b} \Pi_z \sigma_0^{(b)} S_z \, , \label{hbso}
\end{eqnarray}
and they are analogous to the terms in the monolayerlike part
Eq.~(\ref{hmso}).


The monolayerlike [$\mathcal{H}_{m}^{(0)} + \mathcal{H}_{m}^{(1)}$]
and bilayerlike [$\mathcal{H}_{b}^{(0)} + \mathcal{H}_{b}^{(1)}$]
parts of the Hamiltonian produce superimposed monolayerlike $\epsilon_m$
and bilayerlike $\epsilon_b$ bands,
\begin{eqnarray}
\!\!\!\!\!\! \epsilon_m &=& a_m + s \xi \beta_m \pm \sqrt{\left( b_m + s \xi \alpha_m \right)^2 + v^2p^2}
 \, , \label{em} \\
\!\!\!\!\!\! \epsilon_b &=& a_b + s \xi \beta_b \pm \sqrt{\left( b_b + s \xi \alpha_b \right)^2 + v^4p^4/(2\gamma_1^2)} \label{eb} \, ,
\end{eqnarray}
where $s = \pm 1$ denotes different spin components, $\xi = \pm 1$ different valleys,
and we consider $\{ |\epsilon_{m/b}| , vp \} \ll \gamma_1$.
Fig.~\ref{fig:1}(d) shows a plot of the trilayer energy bands,
using Eqs.~(\ref{em}) and (\ref{eb}), where we use $a_m = a_b = b_m = b_b = 0$,
$\alpha_m = \alpha_b = \beta_m = - \beta_b \equiv \alpha$
and the value $\alpha = \gamma_1 / 20$.

The terms proportional to the parameters $a_m$ and $a_b$
in Eqs.~(\ref{hmso}) and (\ref{hbso})
produce constant energy shifts of the monolayerlike
and bilayerlike spectra with respect to each other.
The term proportional to $b_m$ ($b_b$) is absent in a monolayer (bilayer),
but allowed in trilayers owing to the absence of spatial inversion symmetry.
It breaks the $\phi_1^{(m)}$/$\phi_2^{(m)}$ ($\phi_1^{(b)}$/$\phi_2^{(b)}$) sublattice symmetry
and opens a gap between the monolayer (bilayer) bands,
but doesn't generally break valley symmetry \cite{kosh09a}.
The Kane-Mele spin-orbit term, responsible for factors $\alpha_{m/b}$,
has been discussed in detail elsewhere \cite{kanemele05,min06,dhh06,vgeld09,guinea10,liu10}.
Here, we note that, in trilayers, it can conspire with the $b_{m/b}$ parameters
to break spin degeneracy even at zero magnetic field.
Parameters $\beta_{m/b}$ arise from the spin-orbit term unique to odd-$N$
multilayers and they describe a Zeeman-like spin splitting of the energy bands
at each valley, with an opposite sign of the effective Zeeman field in the
two valleys.


The Hamiltonians Eqs. (\ref{hmso}) and (\ref{hbso})
were obtained using
symmetry arguments, but it is possible to relate the phenomenological
parameters to tight-binding parameters. For example, for the spin-independent
part of the Hamiltonian, parameters
$a_m$, $b_m$, $a_b$, $b_b$ may be related to next-nearest-layer coupling
parameters $\gamma_2$ (between $A1$ and $A3$) and $\gamma_5$ (between $B1$ and
$B3$) as \cite{kosh09a} $a_m = - (\gamma_2 + \gamma_5)/2$, $b_m = - (\gamma_2 - \gamma_5)/2$,
$a_b = \gamma_2/2$, $b_b = \gamma_2/2$.


While it is generally accepted that the magnitude $\alpha$
of the intrinsic spin-orbit coupling near the center of the valleys in graphene is small \cite{dresselhaus,kanemele05}, its value
has been the subject of theoretical debate
\cite{min06,dhh06,yao07,boett07,gmitra,guinea10,liu10}.
It was recently proposed \cite{gmitra} that coupling
between the $p_z$ orbitals and $d$ orbitals gives the
dominant contribution to spin-orbit coupling in
monolayers, because $p_z$ orbitals
are not orthogonal to $d_{xz}$ and $d_{yz}$ orbitals,
yielding $\alpha \sim 0.01$meV.
Very recently, it has been suggested \cite{guinea10,liu10} that spin-orbit
coupling in bilayer graphene can be relatively large due to coupling
between $p_z$ orbitals and $\sigma$ bands on different layers (that are
not orthogonal),
giving a contribution to the spin-orbit coupling
$\alpha\sim 0.1$meV.

Using the tight-binding model, we generalize the estimate of
the magnitude of spin-orbit coupling in monolayers \cite{min06,dhh06,yao07}
and bilayers \cite{guinea10,liu10} to demonstrate that {\em both}
types of intrinsic spin-orbit coupling are likely to be as large
in trilayers as the Kane-Mele term in bilayers:
$|\beta_{m/b}| \sim |\alpha_{m/b}|\sim 0.1$meV. We take into account
coupling of the $p_z$ and $\sigma$ orbitals, by writing the tight-binding
Hamiltonian including the Hamiltonian of the $p_z$ orbitals $H_{\pi}$
and of the $\sigma$ orbitals $H_{\sigma}$ as
\begin{eqnarray}
H = \left(
      \begin{array}{cc}
        H_{\pi} & V \\
        V^{\dagger} & H_{\sigma} \\
      \end{array}
    \right) \, , \label{ps}
\end{eqnarray}
where matrix $V$ describes coupling between the $p_z$
and $\sigma$ orbitals.
Here, $H_{\pi}$ is a $12$ by $12$ matrix (describing a spin up and
spin down $p_z$ orbital per site, and six sites in the trilayer unit cell)
and $H_{\sigma}$ is a $36$ by $36$ matrix (six orbitals per site taking
into account two spins and $s$, $p_x$ and $p_y$, and
six sites in the trilayer unit cell).

Block $H_{\sigma}$ contains inter-atomic matrix elements
that are written using Slater and Koster matrix elements \cite{slater}.
For example, at the valley center, 
the matrix element between a $p_x$ orbital on an $Aj$ site
and a $p_y$ orbital on an adjacent $Bj$ site within the same layer
($j=1,2,3$) is $\langle p_x^{Aj} | H | p_y^{Bj} \rangle = - 3 i \xi V_{pp}/4$.
Matrix $V$ in Eq.~(\ref{ps}) describes the coupling between
$p_z$ orbitals and $\sigma$ orbitals.
We take into account skew interlayer coupling between
orbitals on sites $A1$, $B2$ and $A3$,
that are not coupled by $\gamma_1$
(such coupling, if it is only between $p_z$ orbitals,
is usually denoted $\gamma_3$).
For example, the matrix elements between a $p_z$ orbital on the $A1$ site
and a $p_x$ or $p_y$ orbital on the $B2$ site are
$\langle p_z^{A1} | H | p_x^{B2} \rangle = 3 i \xi V_{pp}^{\prime}/2$
and $\langle p_z^{A1} | H | p_y^{B2} \rangle = - 3 V_{pp}^{\prime}/2$
exactly at the valley center. 
For an $A3$ site instead of $A1$, these matrix elements acquire
an additional minus sign.

For the spin-orbit interaction, we include intra-atomic coupling
$H = \lambda_0 {\mathbf{L}}\cdot{\mathbf{S}}$
between the $p$ orbitals. This introduces spin-dependent matrix elements within $H_{\sigma}$
for the same atomic site $j$ such as
$\langle p_{x,\uparrow}^{j} | H | p_{y,\uparrow}^{j} \rangle = - i \lambda_0$,
$\langle p_{x,\downarrow}^{j} | H | p_{y,\downarrow}^{j} \rangle =  i
\lambda_0$.
The effective Hamiltonian describing the $p_z$ orbitals at zero energy may be
written using second-order perturbation theory as
$H_{\rm eff} \approx H_{\pi} - V H_{\sigma}^{-1} V^{\dagger}$.
Doing so, we recover the intrinsic spin-orbit terms, Eqs.~(\ref{hmso}) and (\ref{hbso}), where
$\alpha_m = \alpha_b = \beta_m = - \beta_b = 2\lambda_0 ( V_{pp}^{\prime}/V_{pp})^2$.
The magnitude of this estimate is the same as that for the Kane-Mele
term in bilayer graphene \cite{guinea10}. Using
$V_{pp}^{\prime}/V_{pp} \sim 0.1$ and $\lambda_0 \sim 10$meV
yields $|\alpha_{m/b}| = |\beta_{m/b}| \sim 0.1$meV.
We compare the spin splitting of the bands, $2|\beta_{m/b}|$, to the Zeeman
energy $g \mu_B B$ in a real external field $B$. Using $g = 2$
gives an effective field of $B \sim 2$T,
which is comparable to the real fields
required to observe Zeeman-split conductance fluctuations
in graphene \cite{lund09}.

The authors thank V. I. Fal'ko and H. Schomerus for discussions.
This project has been funded by EPSRC First Grant EP/E063519/1 and by
JST-EPSRC Japan-UK Cooperative Programme Grant EP/H025804/1.

\end{document}